\documentstyle[prl,aps,psfig]{revtex}

\begin{document}
\twocolumn
\twocolumn[\hsize\textwidth\columnwidth\hsize\csname
@twocolumnfalse\endcsname

\title {Angle-resolved photoemission study of 
insulating and metallic Cu-O chains
in PrBa$_2$Cu$_3$O$_7$ and PrBa$_2$Cu$_4$O$_8$} 
\author{T.~Mizokawa$^{1,4}$, C.~Kim$^{2,3}$, Z.-X.~Shen$^{2,3}$,
A.~Ino$^4$, T.~Yoshida$^4$, A.~Fujimori$^{1,4}$, 
M.~Goto$^5$, H.~Eisaki$^{2,5}$, S.~Uchida$^{5,6}$, M.~Tagami$^7$, 
K.~Yoshida$^7$, A.~I.~Rykov$^7$, Y.~Siohara$^7$, K.~Tomimoto$^7$,
S.~Tajima$^7$, Yuh~Yamada$^8$, S.~Horii$^6$, N.~Yamada$^9$,
Yasuji~Yamada$^7$, I.~Hirabayashi$^7$
} 
\address{$^1$ Department of Complexity Science and Engineering, 
University of Tokyo, Bunkyo-ku, Tokyo 113-0033, Japan}
\address{$^2$ Department of Applied Physics, Stanford University, 
Stanford, CA94305 , U.S.A.}
\address{$^3$ Stanford Synchrotron Radiation Laboratory, 
Stanford University, Stanford, CA94305 , U.S.A.}
\address{$^4$ Department of Physics, University of Tokyo, Bunkyo-ku, 
Tokyo 113-0033, Japan}
\address{$^5$ Department of Superconductivity, University of Tokyo, 
Bunkyo-ku, Tokyo 113-0033, Japan}
\address{$^6$ Department of Advanced Materials, University of Tokyo, 
Bunkyo-ku, Tokyo 113-0033, Japan}
\address{$^7$ Superconductivity Research Laboratory,
International Superconductivity Technology Center, Koto-ku, Tokyo
135-0062, Japan}
\address{$^8$ Faculty of Science and Engineering, Shimane University, 
Matsue 690-0823, Japan}
\address{$^9$ Department of Applied Physics and Chemistry, University of
Electro-Communications, Chofu, Tokyo 182-8585, Japan}
\date{\today}
\maketitle

\begin{abstract}
We compare the angle-resolved photoemission spectra of the hole-doped Cu-O chains 
in PrBa$_2$Cu$_3$O$_7$ (Pr123) and in PrBa$_2$Cu$_4$O$_8$ (Pr124).
While, in Pr123, a dispersive feature from the chain
takes a band maximum at $k_b$ (momentum along the chain)
$\sim$ $\pi/4$ and loses its spectral weight around the Fermi level,
it reaches the Fermi level at $k_b$ $\sim$ $\pi/4$ in Pr124. 
Although the chains in Pr123 and Pr124 are approximately 
1/4-filled, they show contrasting behaviors:
While the chains in Pr123 have an instability to charge ordering,
those in Pr124 avoid it and show an interesting spectral feature 
of a metallic coupled-chain system.
\end{abstract}

\pacs{PACS numbers: 74.25.Jb, 71.27.+a, 73.20.Dx, 79.60.Bm}

]


One-dimensional (1D) or quasi 1D compounds have attracted
great interest because of their unique and interesting physical properties
such as Peierls instability, spin-charge separation, and Tomonaga-Luttinger (TL) liquid behavior \cite{1D}.
In order to realize the TL liquid state, 
the system should have good one dimensionality.
On the other hand, a good 1D system is expected to have
instability towards spin and/or charge ordering and tends to be insulating.
In every 1D system, a metallic state which is possibly a TL liquid,
is competing with the instability to the insulating state. 
It is therefore very important and interesting to study 
the electronic structure of quasi 1D systems which are 
close to the boundary between the quasi 1D metal and insulator.
This subject is also related to the electronic structure
of the quasi 1D stripe phase in 2D systems 
and is important in the light of the competition 
between various quantum mechanical ground states
in 1D and 2D systems \cite{Sachdev}.

Angle-resolved photoemission spectroscopy (ARPES) is a powerful tool
to study spin-charge separation as well as band gap opening
due to charge ordering. It has been found that
ARPES of the Cu-O double chain in SrCuO$_2$
and the Cu-O single chain in Sr$_2$CuO$_3$
have two dispersive features which are successfully
interpreted as spinon and holon dispersions 
of the undoped Cu-O chains and are manifestation of 
the spin-charge separation in 1D Mott insulators \cite{Kim}.
In an ARPES study of PrBa$_2$Cu$_3$O$_7$ (Pr123), it has been reported that 
the hole-doped Cu-O single chain in Pr123 shows possible spinon and holon
dispersions and also has band gap opening at Fermi level ($E_F$)
probably due to charge ordering \cite{Mizokawa}.
Recently, an ARPES study of a quasi 1D metal Li$_{0.9}$Mo$_6$O$_{17}$
has shown that spectral weight near $E_F$ is considerably suppressed
and can be interpreted as a TL liquid with large $\alpha$
which is the anomalous exponent of the momentum distribution function 
as well as of the single-particle density of states \cite{Allen}. 
The large $\alpha$ suggests strong fluctuations
of charge ordering or charge density wave (CDW) in Li$_{0.9}$Mo$_6$O$_{17}$
although the analysis of the ARPES data is still controversial
\cite{Allen2}.

In Pr123 and PrBa$_2$Cu$_4$O$_8$ (Pr124), 
the CuO$_2$ planes remain antiferromagntic and insulating
and do not have enough carriers to cause superconductivity
\cite{Pr123,Takenaka,Pr124}.
On the other hand, the Cu-O single chains in Pr123
and the Cu-O double chains in Pr124 are heavily hole-doped and show 
semiconducting and metallic behaviors, respectively 
\cite{Pr123,Takenaka,Pr124}. 
Therefore, Pr123 and Pr124 give us a unique opportunity to study
the electronic structures of the hole-doped Cu-O chains
which are close to the boundary between the quasi 1D
metal and insulator. In this Letter, we report on a new set of 
ARPES data for Pr124 which has metallic Cu-O double chains.
The ARPES data of the Cu-O double
chain in Pr124 shows a sharp contrast to
those of the Cu-O single chain in Pr123.
Since the structures of the Cu-O chains in Pr123 and Pr124,
which are the same as the Cu-O chains in Sr$_2$CuO$_3$ and SrCuO$_2$
respectively, are simple compared to Li$_{0.9}$Mo$_6$O$_{17}$,
the interpretation of the ARPES data is rather straightforward.
By comparing the ARPES data of Pr123 and Pr124,
we discuss the effect of interchain coupling and the instability 
towards charge ordering in the hole-doped Cu-O chains.


Naturally-untwinned single crystals of Pr124 were grown 
by a flux method under oxygen pressure of 11 atm.
The resistivity along the chain direction
is metallic as reported in the literature \cite{Terasaki,Horii}.
The ARPES measurements of Pr124 were performed at beamline 
5-4 equipped with a Scienta SES 200 electron analyzer,
Stanford Synchrotron Radiation Laboratory (SSRL).
The chamber pressure during the measurements was less than
5 x 10$^{-11}$ Torr.  The samples were cooled to
10 K and cleaved {\it in situ}. The cleaved surfaces
were the $ab$-plane, where the $b$-axis is in the Cu-O chain
direction. The cleanliness of the surfaces was checked by the absence
of a hump at $\sim$ 9.5 eV. The position of $E_F$ was
calibrated with gold spectra. The experimental uncertainty
in the energy calibration was $\pm$ 1 meV.
The ARPES data of Pr123 were taken
at beamlines 5-3 of SSRL. The details of the measurements
were described in the previous paper \cite{Mizokawa}.
For the ARPES data of Pr123 and Pr124 shown in this paper,
incident photons were linearly polarized and had an energy of
29 eV for Pr123 and 22.4 eV for Pr124.
The total energy resolution including the 
monochromator and the analyzer was approximately 40 meV
for Pr123 and 20 meV for Pr124.
The angular resolution was $\pm 1$ degree for Pr123
and $\pm 0.28$ degree for Pr124, which gives 
the momentum resolution of $\pm$ 0.05$\pi$ for Pr123 and 
$\pm$ 0.01$\pi$ for Pr124 in units of $1/a$ or $1/b$. 
($a$ = 3.87 ${\rm \AA}$ and $b$ = 3.93 ${\rm \AA}$ for Pr123, 
$a$ = 3.88 ${\rm \AA}$ and $b$ = 3.90 ${\rm \AA}$ for Pr124)  

\begin{figure}
\psfig{figure=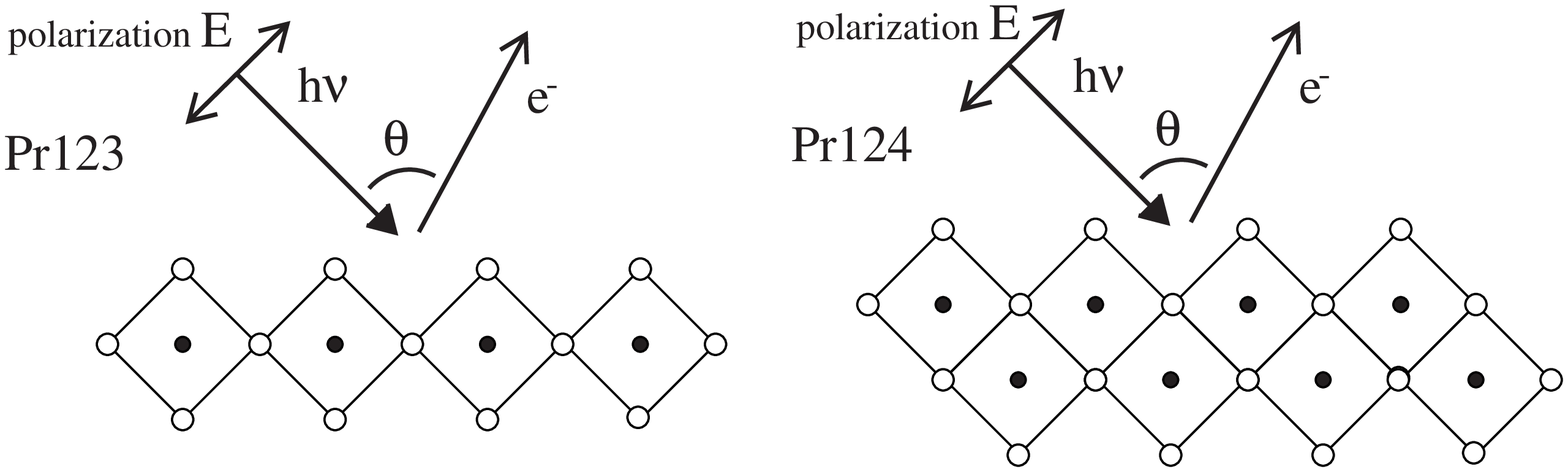,width=9cm}
\caption{
Schematic drawings of the experimental arrangement
and the Cu-O single and double chains
at the cleaved surfaces of Pr123 and Pr124. 
The closed and open circles indicate Cu and oxygen 
ions, respectively.
}
\label{structure}
\end{figure}
\begin{figure}
\psfig{figure=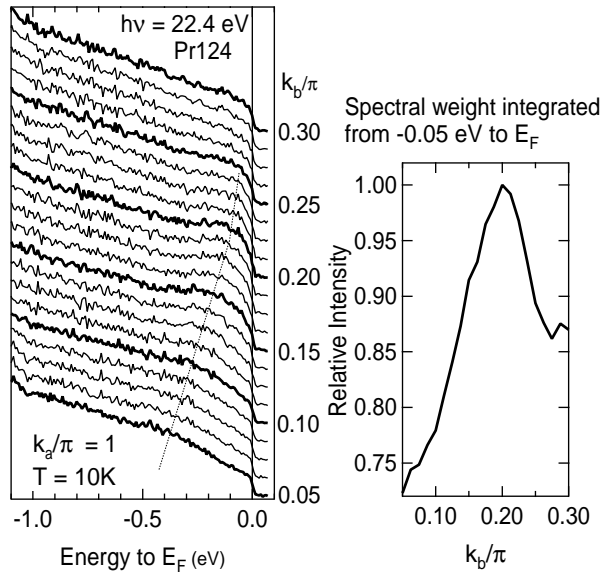,width=9cm}
\caption{
Left panel: ARPES spectra along the Cu-O chain in Pr124.
$k_a$ is the momemtum perpendicular to the chain
and $k_b$ is the momentum along the chain. 
The dotted line outlines the band dispersion.
Right panel: Spectral weight integrated from -0.05 eV
to $E_F$ as a function of the momentum along the chain.
Intensity is normalized to the peak height.
}
\label{ARPESPr124}
\end{figure}

The experimental arrangement is schematically shown
in Fig. \ref{structure}. The polarization vector of the incident photons 
had a component parallel to the Cu-O chain direction. 
The CuO$_4$ square planes of the Cu-O chains 
are perpendicular to the cleaved surface.
In Fig. \ref{ARPESPr124}, the ARPES spectra of Pr124
along the Cu-O chain direction are shown 
for $k_a = \pi$ \cite{TME}. 
Here, $k_a$ and $k_b$ are the momentum perpendicular
to the chain in units of $1/a$ 
and the momentum along the chain
in units of $1/b$, respectively.  
In Fig. \ref{ARPESPr124}, one can see that
a dispersive feature from the Cu-O chain
moves to $E_F$ in going from $k_{b}/\pi = 0.1$ to
$k_{b}/\pi = 0.2$. For $0.2 < k_{b}/\pi < 0.25$,
the dispersion becomes flatter and the dispersive 
feature gradually loses its intensity.
This behavior is clearly seen in the right panel
of Fig. \ref{ARPESPr124}, where the spectral weight 
integrated from -0.05 eV to $E_F$ is plotted 
as a function of $k_{b}/\pi$. The integrated
spectral weight decreases from $k_{b}/\pi = 0.20$ to 0.27,
indicating that the dispersive feature crosses
$E_F$ at $k_{b}/\pi = 0.23 \pm 0.03$ \cite{ebar}
and that the hole concentration of the Cu-O chain
in Pr124 is 0.46 $\pm$ 0.06. This is consistent with 
the recent optical study which shows that the hole concentration
of the Cu-O chain in Pr124 is $\sim$ 0.4 \cite{Takenaka2}.

\begin{figure}
\psfig{figure=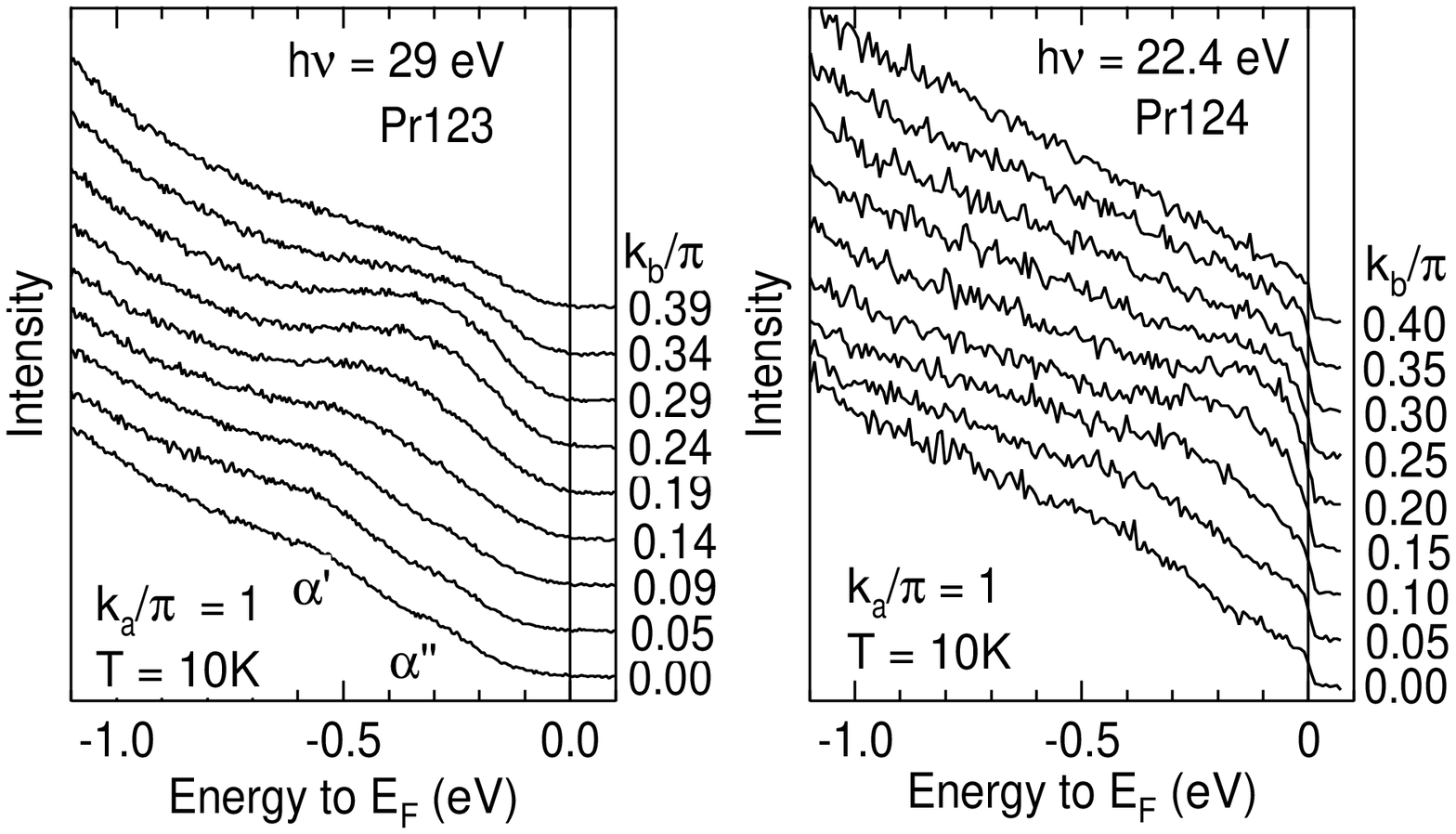,width=8.5cm}
\caption{
ARPES spectra taken along the Cu-O chain in Pr123 and Pr124.
$k_a$ is the momemtum perpendicular to the chain
and $k_b$ is the momentum along the chain.
}
\label{ARPES}
\end{figure}

\begin{figure}
\psfig{figure=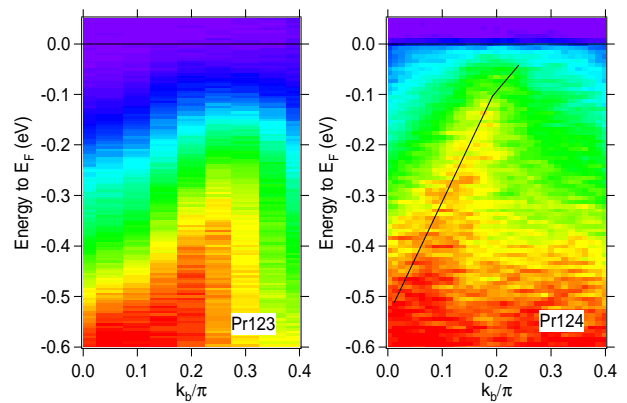,width=8.5cm}
\caption{
Density plot of the ARPES spectra for Pr123 and Pr124
along the chain direction. $k_b$ is the momemtum along the chain. 
Intensity increases in going from blue to red regions.
The solid line outlines the Cu-O band dispersion in Pr124.
While, in Pr124, the width of the dispersion below $E_F$
is $\sim$ 0.5 eV, the band width is reduced to $\sim$ 0.3 eV
in Pr123 because of the band gap opening.
}
\label{density}
\end{figure}

The ARPES spectra of Pr123 and Pr124 along the Cu-O chain
are compared in Fig. \ref{ARPES} for $k_a = \pi$. 
In Pr123, the dispersive feature from the Cu-O chain reaches the band
maximum of -0.3 eV at $k_{b}/\pi = 0.24$ and stays there at $k_{b}/\pi = 0.29$.
This feature gradually loses its weight for $k_{b}/\pi > 0.29$
without reaching $E_F$.
On the other hand, in Pr124, the dispersive
feature reaches $E_F$ at $k_{b}/\pi \sim 0.23 \pm 0.03$.
Although the Cu-O chains are approximately 1/4-filled
both in Pr123 and in Pr124, the spectral feature near
$E_F$ of Pr124 is remarkably different from that of Pr123.
As shown in Fig. \ref{density}, 
the dispersions of the Cu-O chain features become clear
in the density plots of the raw ARPES spectra.
While, in Pr124, the width of the Cu-O band below $E_F$
is $\sim$ 0.5 eV and the band dispersion is relatively large,
the band width is reduced to $\sim$ 0.3 eV in Pr123
because of the band gap opening at $E_F$.

For Pr123, the band gap opening can be attributed 
to the charge ordering or CDW formation
in the 1/4-filled Cu-O single chain \cite{Mizokawa}. 
Actually, charge instability in the Cu-O chain 
has been observed by NMR and NQR measurements of Pr123 \cite{NMR}.
The CDW picture is also consistent with the experimental result
that the gap opening at $E_F$ reduces the band width of Pr123 
compared to Pr124. On the other hand, the Cu-O double chain of Pr124
does not have charge ordering and is metallic.
A small deviation from the 1/4-filling would be
responsible for this suppression of charge ordering in Pr124.
It is also possible that, since each double 
chain of Pr124 consists of two single chains
(see Fig. \ref{structure}), charge ordering
is unstable because of the weak interaction
between the two single chains in each double chain \cite{MizokawaHF}.

The ARPES study of Li$_{0.9}$Mo$_6$O$_{17}$
shows that the spectral weight near $E_F$ is considerably 
suppressed even in the metallic region and gives the
anomalous exponent $\alpha$ of 0.9 \cite{Allen},
indicating that Li$_{0.9}$Mo$_6$O$_{17}$ is a TL liquid 
close to the CDW instability.
On the other hand, the ARPES data of Pr124
have substantial spectral weight at $E_F$ compared 
to Li$_{0.9}$Mo$_6$O$_{17}$ \cite{Allen}
and, in the low-energy region, Pr124 does not
show the TL-liquid behavior. 
This is consistent with the fact that Pr124
has large Hall coefficient at low temperature 
and behaves as a 2D system \cite{Terasaki}. 
Probably, the deviation from the TL-liquid behavior
is caused by the hopping between the double chains
which becomes relevant in the low-energy region.

\begin{figure}
\psfig{figure=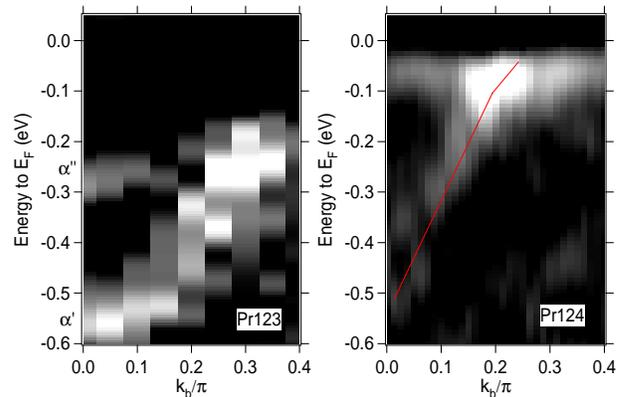,width=8.5cm}
\caption{
Second derivatives of the ARPES spectra for Pr123 and Pr124
along the chain direction. $k_b$ is the momemtum along the chain. 
The red solid line outlines the Cu-O band dispersion in Pr124.
In Pr123, two dispersive features are visible as two bright belts.
}
\label{2nd}
\end{figure}

The hopping term between the double chains is estimated 
to be $\sim$ 10 meV from the 1D-2D crossover observed 
in the transport measurements of Pr124,\cite{Terasaki}.
In the energy region higher than the hopping term,
the 1D character is expected to manifest in the ARPES spectra.
In order to show the dispersions more clearly, the second derivatives of
the ARPES spectra are displayed in Fig. \ref{2nd}. In Pr123, two dispersive 
features labeled as $\alpha'$ and $\alpha''$ are visible as two bright belts 
which can be attributed to holon and spinon dispersions, respectively, and are
the manifestations of the spin-charge separation in 1D systems 
as predicted theoretically \cite{TL}.
The holon and spinon dispersions have the width 
of $\sim$ 0.5 and $\sim$ 0.1 eV, respectively, which
approximately agree with $t$ and $J$ in the $t$-$J$ model
for the cuprates \cite{Maekawa}.
On the other hand, Pr124 does not show separate spinon and holon 
features expected for the 1D $t$-$J$ model \cite{Maekawa}. 
One possible explanation
is that, in the high-energy region, Pr124 behaves as a TL-liquid
with anomalous exponent $\alpha$ larger than 0.5
because of proximity to charge ordering as proposed 
in the recent optical study on Pr124 \cite{Takenaka2}.
In this case, since the TL model predicts that 
the spinon feature appears as a cusp 
instead of a power-law divergence, 
the spinon feature is expected to be broad and 
may not separately be observed from the holon feature 
\cite{Allen,TL}.

As seen in Figs. \ref{ARPES} and \ref{density}.
the ARPES spectra for $k_b/\pi$ $<$ 0.20 have
substantial spectral weight near $E_F$ although the dispersive feature
is located well below $E_F$ (see Fig. \ref{ARPESPr124}).
In the second derivative shown in Fig. \ref{2nd}, this spectral weight 
near $E_F$ appears as a horizontal bright belt near $E_F$
ranging from $k_b/\pi$ = 0.0 to 0.4.
In addition, for $0.20 < k_b/\pi < 0.25$, the dispersion becomes flatter
and it looks like that a small pseudo-gap tend to open at $E_F$. 
These behaviors in the metallic Cu-O chains of Pr124 
are similar to those observed along $(\pi,0) \rightarrow (\pi,\pi)$ 
above $T_{c}$ in the underdoped CuO$_2$ plane of the high-$T_{c}$
cuprates \cite{Bi2212}, which in the stripe phase
represent the dispersion along the 1/4-filled stripe \cite{Zou}.
The similarity between the coupled chains and the high-$T_c$ cuprates 
has been argued by Kopietz, Meden, and Sch\"{o}nhammer 
in the light of the crossover between a TL liquid 
and a 2D Fermi liquid \cite{Kopietz}. It would be 
interesting to investigate whether the coupled-chain model 
can explain the present ARPES data of Pr124 as well as
those of the high-$T_c$ cuprates.

In conclusion, we have studied the hole-doped Cu-O chains in Pr123 and Pr124
using ARPES. The ARPES data show that the Cu-O single chain in Pr123
and the Cu-O double chain in Pr124 are approximately 1/4-filled.
While, in Pr123, the 1D features from the Cu-O single chain 
lose their weight near $E_F$, the dispersive feature
from the Cu-O double chain reaches $E_F$ in Pr124.
These facts indicate that the charge ordering occurs in the Cu-O single
chain and is suppressed in the Cu-O double chain. In the low-energy region,
the line shape of the ARPES spectra of Pr124
cannot be explained neither by the TL-liquid picture nor
by the Fermi-liquid picture and looks rather close to that of
the 2D CuO$_2$ plane with stripes. 
It would be useful to further study the differences and similarities
between the metallic Cu-O chains in Pr124 and the metallic stripe phase
in the high-$T_c$ cuprates.


The authors would like to thank K. Takenaka,
I. Terasaki, K. Penc, T. Thoyama, and S. Maekawa 
for valuable comments and the staff of SSRL for technical support.
This work is supported by a Grant-in-Aid for Scientific Research
"Novel Quantum Phenomena in Transition Metal Oxides"
from the Ministry of Education, Science, Sports and Culture of Japan,
Special Coordination Funds of the Science and Technology Agency of Japan,
the New Energy and Industrial Technology Development
Organization (NEDO), the U.~S.~DOE, Office of Basic Energy 
Science and Division of Material Science. SSRL 
is operated by the U.~S.~DOE, Office of Basic Energy 
Sciences, Division of Chemical Sciences.

\begin{references}
\bibitem{TME}
The intensity of the Cu 3$d_{x^2-y^2}$-O 2$p_{\sigma}$ 
derived band in the Cu-O chain becomes strong as $k_a$ increases 
and the angle $\theta$ between the Poynting vector
of the incoming photons and the direction of the outgoing electrons
becomes small. This momentum dependence of the
intensity is qualitatively explained by 
the transition matrix element of the $d_{x^2-y^2}$-like 
orbitals in the Cu-O chains.
\bibitem{ebar}
We put the large error bar because the dispersive feature
does not show a typical Fermi-liquid behavior and, consequently,
it is very difficult to precisely estimate 
the momentum at which it crosses the Fermi level.
\end{references}
\end{document}